\documentclass[prb,twocolumn,showpacs,aps,superscriptaddress,floatfix]{revtex4}
\usepackage{amsmath}
\usepackage{amssymb}
\usepackage{bm}
\usepackage{graphicx}
\usepackage{color}
\usepackage{hyperref}
\usepackage{verbatim}

\DeclareMathOperator{\arsinh}{arsinh}

\begin{document}

\title{Spin-density wave state in simple hexagonal graphite}

\author{K.S. Mosoyan}
\affiliation{Moscow Institute for Physics and Technology (State
University), Moscow region, 141700 Russia}

\author{A.V. Rozhkov}
\affiliation{Moscow Institute for Physics and Technology (State
University), Moscow region, 141700 Russia}
\affiliation{Institute for Theoretical and Applied Electrodynamics, Russian
Academy of Sciences, Moscow, 125412 Russia}

\author{A.O. Sboychakov}
\affiliation{Institute for Theoretical and Applied Electrodynamics, Russian
Academy of Sciences, Moscow, 125412 Russia}

\author{A.L. Rakhmanov}
\affiliation{Moscow Institute for Physics and Technology (State
University), Moscow region, 141700 Russia}
\affiliation{Institute for Theoretical and Applied Electrodynamics, Russian
Academy of Sciences, Moscow, 125412 Russia}
\affiliation{Dukhov Research Institute of Automatics, Moscow, 127055
Russia}

\begin{abstract}
Simple hexagonal graphite, also known as AA~graphite, is a metastable
configuration of graphite. Using tight-binding approximation it is easy to
show that AA~graphite is a metal with well-defined Fermi  surface. The
Fermi surface consists of two sheets, each shaped like a rugby ball.  One
sheet corresponds to electron states, another corresponds to hole states.
The Fermi surface demonstrates good nesting: a suitable translation in the
reciprocal space superposes one sheet onto another. In the presence of the
electron-electron repulsion a nested Fermi surface is unstable with respect
to spin-density wave ordering. This instability is studied using the
mean-field theory at zero temperature, and the spin-density wave order
parameter is evaluated.
\end{abstract}

\date{\today}

\maketitle

\section{Introduction}

Since recent isolation of the graphene
layer~\cite{graphene_novos2004}
the interest to layered carbon systems was reignited.
It has been known for some time already that such systems are very diverse,
and demonstrate interesting many-body electron properties. For example,
graphite in magnetic field
undergoes~\cite{cdw_magfield_exp2017}
a transition into a field-induced charge-density wave (CDW) state. After
intercalation graphite may become a superconductor. For
example~\cite{supercond_inter2005},
the critical temperature for graphite intercalated with Ca equals to
$T_c=11.5$\,K,
as for Yb-intercalated graphite, it is characterized by
$T_c=6.5$\,K.

\begin{figure}[t]
\center{\includegraphics[width=0.95\columnwidth]{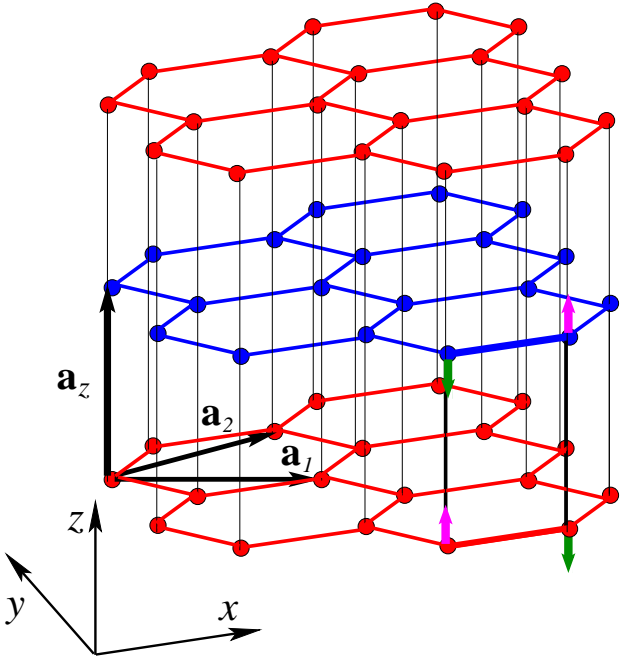}}
\caption{Simple hexagonal lattice of AA~graphite. It consists of layers of
graphene stacked upon each other. Lattice vectors are
$\mathbf{a}_1$,
$\mathbf{a}_2$,
and
$\mathbf{a}_z$.
The distance between neighboring atoms
inside the layer is
$a=1.42$\,\AA, while the distance between the layers is
$c\equiv|{\bf a}_z|\approx3.3$\,\AA\ according to
Ref.~\onlinecite{aa_graphite_charlier1994} and
$c\approx 3.4$\,\AA\
according to Ref.~\onlinecite{borysiuk2011_aa}. Elementary unit cell of the
lattice of AA~graphite consists of two atoms corresponding to two
non-equivalent sublattices of the graphene layer. The SDW ordering doubles
the lattice period in the
$z$-direction,
making spins configurations in neighboring graphene layers different from
each other. As a result, the magnetic unit cell contains four atoms. Short
thick arrows show the spin configuration inside this unit
cell. 
\label{ris::aa_latt}}
\end{figure}

In this paper a purely carbon system, simple hexagonal graphite [also known
as AA~graphite (AA-G)] is discussed. A fragment of simple hexagonal lattice
is shown in
Fig.~\ref{ris::aa_latt}.
It is believed~\cite{aa_graphite_charlier1994}
that the simple hexagonal lattice has higher energy than the hexagonal
(also referred to as ABA) and rhombohedral (ABC) lattices. In other words,
among the three possible highly-symmetric layered structures of carbon, the
simple hexagonal lattice is the least stable. This implies that
experimental realization of the AA-G is bound to run into difficulties:
AA~lattice will try to relax into either ABA or ABC~structures to reduce
the chemical energy. Yet, samples of AA-G (as well as bilayer and
multi-layer AA~graphene, which are similar to the AA-G) were synthesized by
several
groups~\cite{aa_graphite_lee2008,liu_aa_exp2009,borysiuk2011_aa,
aa_graphit_roy1998}.
These experimental advances make the studies of electron properties of the
AA-G a timely theoretical task.

From the band theory standpoint, the AA-G is a metal with a well-defined
Fermi
surface~\cite{aa_graphite_charlier1994,aa_tb_charlier1991,
aa_abinit_charlier1992}.
The Fermi surface consists of two sheets, or two components. One component
corresponds to electron states, the other component corresponds to hole
states. Both sheets have shapes of rugby balls. The sheet shapes are
almost identical, and suitable translation superposes them. The latter
property of the Fermi surface is called nesting.

A Fermi surface with the nesting is unstable with respect to the spin
density wave (SDW) order. The instability is driven by electron-electron
repulsion. The main purpose of this paper is to discuss the SDW instability
of the AA-G electronic liquid at zero temperature. Using tight-binding
approximation we will evaluate the Fermi surface structure of the AA-G, and
demonstrate that the nesting of the Fermi surface is indeed present. After
that, the SDW zero-temperature state will be studied with the help of
mean-field approximation.

The paper is organized as follows. In
Sec.~\ref{sec::tb}
we formulate the tight-binding description of the AA-G. The
zero-temperature mean-field calculations are performed in
Sec.~\ref{sec::mean_f}.
Finally,
Sec.~\ref{sec::conclusions}
presents both the discussion and the conclusions of the study. Technically
involved details are relegated to Appendices.

\section{Tight-binding model of the AA~graphite}
\label{sec::tb}

\subsection{Geometry and tight-binding description of graphene}

Tight-binding model of the AA-G is a straightforward generalization of the
tight-binding model of graphene. The latter is mostly determined by the
geometrical properties of the honeycomb lattice of graphene (for more
details one can consult a review on graphene, for example,
Ref.~\onlinecite{meso_review}). 
The graphene has hexagonal lattice consisting of two triangular
sublattices, $A$ and $B$
(see
Fig.~\ref{ris::aa_latt}).
Thus, elementary unit cell of graphene contains two atoms. The elementary
translation vectors may be chosen as follows
\begin{eqnarray}\mathbf{a}_1 = \dfrac{a}{2} \left( 3,-\sqrt3\right),\qquad\mathbf{a}_2 = \dfrac{a}{2} \left( 3,\sqrt3 \right),
\end{eqnarray}
where
$a\approx1.42$\AA\
is the distance between the nearest-neighbor carbon atoms. The reciprocal
lattice vectors are
\begin{eqnarray}\mathbf{b}_1 = \dfrac{2\pi}{3a}\left( 1,-\sqrt3\right),\qquad\mathbf{b}_2 = \dfrac{2\pi}{3a}\left(1,\sqrt3 \right).
\end{eqnarray}
The Dirac cones of the graphene are located in the corners of the hexagonal
Brillouin zone. Without loss of generality we can assume that these cones
are centered at points
\begin{eqnarray}\label{eq::K}
\mathbf{K}= \left(\dfrac{2\pi}{3a},\dfrac{2\pi}{3\sqrt{3}a} \right),\qquad\mathbf{K^\prime} = \left(\dfrac{2\pi}{3a},-\dfrac{2\pi}{3\sqrt{3}a} \right).
\end{eqnarray}
For the single layer (thus, abbreviation `sl') of graphene the simplest
tight-binding Hamiltonian for
$\pi$-bonds 
of carbon atoms equals
\begin{equation}\label{Hsl}
H^{\rm sl}
=
-t\!\!
\sum_{\langle\mathbf{nm}\rangle\sigma }\!\!\left(d_{\mathbf{n}A\sigma }^{\dag}d_{\mathbf{m}B \sigma }^{\phantom{\dag}}+\text{h.c.}\right).
\end{equation}
Here
$d_{\mathbf{n}\alpha\sigma}^{\dag}$
and
$d_{\mathbf{n}\alpha\sigma}^{\phantom{\dag}}$
are the creation and annihilation operators of the electron with spin
projection $\sigma$, located at the unit cell
$\mathbf{n}=(n,m)$
($n$ and $m$ are integers) in the sublattice
$\alpha=A,\,B$.
The summation in
Eq.~\eqref{Hsl}
is performed over nearest neighbor sites, and
$t\approx2.7$\,eV
is the nearest-neighbor hopping integral. We introduce the Fourier
transformed electronic operators
$d_{\mathbf{k}\alpha\sigma}^{\phantom{\dag}}
=
\sum_{\mathbf{n}}
	e^{i\mathbf{kr}_{\mathbf{n}}^{\alpha}}
	d_{\mathbf{n}\alpha\sigma}^{\phantom{\dag}}/\sqrt{\cal N}$,
where
$\mathbf{r}_{\mathbf{n}}^{\alpha}$
is the position of a carbon atom in the
$\mathbf{n}$-th
unit cell for sublattice $\alpha$, while
${\cal N}$
is the number of unit cells in the sample. We also define the
(pseudo)spinor
\begin{eqnarray}\label{eq::Psi}
\psi_{\mathbf{k} \sigma}^{\phantom{\dag}}=
\begin{pmatrix}
d_{\mathbf{k}A\sigma}^{\phantom{\dag}}\\
d_{\mathbf{k}B\sigma}^{\phantom{\dag}}
\end{pmatrix}.
\end{eqnarray}
The Hamiltonian~\eqref{Hsl} can be rewritten as
\begin{eqnarray}\label{eq::h_grEn}
H_{\rm sl}=\sum\limits_{\mathbf{k}\sigma}\psi_{\mathbf{k}\sigma}^{\dag}\hat{H}_{\mathbf{k}}^{\rm sl}\psi_{\mathbf{k}\sigma}^{\phantom{\dag}}.
\end{eqnarray}
where
$2\times2$
matrix
$\hat{H}_{\mathbf{k}}^{\rm sl}$
is
\begin{eqnarray}
\hat{H}_{\mathbf{k}}^{\rm sl}=-t\begin{pmatrix}
0&f(\mathbf{k})\\f^*(\mathbf{k})&0\end{pmatrix}.
\end{eqnarray}
In this expression function $f$ is equal to
\begin{equation}
f(\mathbf{k})=e^{-iak_x}\left[1+2e^{3iak_x/2}\cos\left(\frac{\sqrt{3}}{2}ak_y\right)\right].
\end{equation}

For a given value of quasimomentum
${\bf k}$
the eigenvalues of
Eq.~(\ref{eq::h_grEn}) are equal to
$\varepsilon_{\mathbf{k}}^{(1,2)}=\pm t|f(\mathbf{k})|$.
Near the Brillouin zone corners function
$f(\mathbf{k})$
can be expanded as
\begin{eqnarray}
\label{eq::f_exp1}
f(\mathbf{K}+\mathbf{q})
\approx
\frac{3a}{2} e^{-\frac{2\pi i}{3}}\left(q_y-iq_x\right),\;
\\
\label{eq::f_exp2}
f(\mathbf{K}'+\mathbf{q})
\approx
\frac{3a}{2} e^{-\frac{2\pi i}{3}}\left(-q_y-iq_x\right).
\end{eqnarray}
If we substitute
Eq.~\eqref{eq::f_exp1}
and~\eqref{eq::f_exp2}
into
Hamiltonian~\eqref{eq::h_grEn},
the latter becomes equivalent to two two-dimensional (2D) Dirac-Weyl
Hamiltonians of massless relativistic fermions. Their dispersion is
\begin{equation}
\varepsilon_{\textbf{q}}^{(1,2)}=\pm  v_{\rm F}|\textbf{q}|\,.
\end{equation}
Here the Fermi velocity
$v_{\rm F}=3at/2$
plays the role of speed of light.

\subsection{Tight-binding description of the AA~graphite}

Hamiltonian~(\ref{eq::h_grEn})
can be easily modified to describe AA~graphite. The generalized
Hamiltonian should account for a macroscopic number of stacked graphene
layers coupled by single-electron hopping. Electrons with different spins
are decoupled from each other. Consequently, we can write
\begin{equation}
\label{eq::HAAtbfull}
H^{\rm AA}=\sum_\sigma H_\sigma^{\rm AA}\,,
\end{equation}
where
\begin{eqnarray}
\label{eq::HAAtb}
H_\sigma^{\rm AA}
&=&-t\!\!
\sum_{\langle\mathbf{nm}\rangle i}\!\!
	\left(
		d_{\mathbf{n}iA\sigma}^{\dag}
		d_{\mathbf{m}iB\sigma }^{\phantom{\dag}}
		+
		\text{h.c.}
	\right)
\nonumber\\
&&-t_0
\sum_{\mathbf{n}i\alpha}
	\left(
		d_{\mathbf{n}i+1\alpha\sigma}^{\dag}
		d_{\mathbf{n}i\alpha\sigma }^{\phantom{\dag}}
		+
		\text{h.c.}
	\right).
\end{eqnarray}
In this expression integer $i$ enumerates the layers. The first sum
describes the in-layer electron hopping, while the second sum corresponds
to the nearest-neighbor inter-layer hopping. The inter-layer hopping
amplitude
$t_0$
is about
$0.3$\,--\,$0.4$\,eV.

\begin{figure}[t]
\center{\includegraphics[width=0.95\columnwidth]{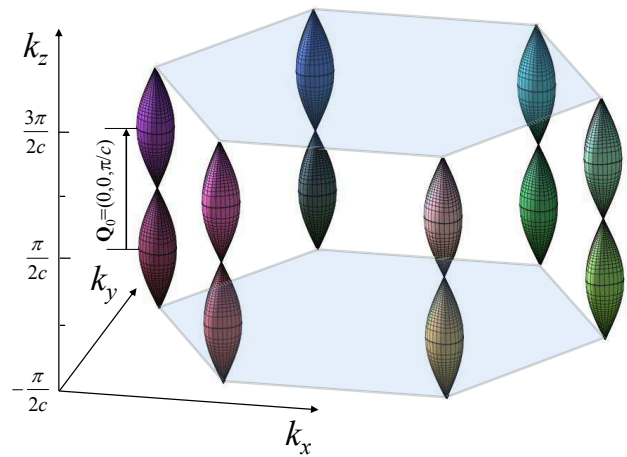}}
\caption{Fermi surface of the AA-G inside the first Brillouin zone
calculated for
$t=2.7$\,eV
and
$t_0=0.4$\,eV.
The Brillouin zone is shifted by
$-\pi/(2c)$
along
$z$-axis for clarity. The Fermi surface emerges near each corner of the
Brillouin zone. The Fermi surface consists of two sheets with a shape of a
rugby ball. The lower sheet is electron-like, while the upper one is the
hole-like. Two sheets coincide upon the translation by the nesting vector
$\mathbf{Q}_0=(0,\,0,\,\pi/c)$.
\label{FigFermiSurface}}
\end{figure}

Elementary unit cell of the AA-G contains two atoms and is characterized by
vectors
$\mathbf{a}_1$,
$\mathbf{a}_2$,
and
$\mathbf{a}_{z}=c\,\mathbf{e}_z$,
where
$\mathbf{e}_z$
is the unit vector along
$z$-axis
perpendicular to the layers, while
$c\approx3.3$\,\AA\
is the inter-layer distance. Reciprocal unit cell of the AA-G  is
characterized by vectors
$\mathbf{b}_1$,
$\mathbf{b}_2$,
and
$\mathbf{b}_{z}=2\pi\mathbf{e}_z/c$.
We introduce Fourier transformed operators
\begin{equation}
d_{\mathbf{k}\alpha\sigma}^{\phantom{\dag}}=\frac{1}{\sqrt{\cal N}}\sum_{\mathbf{n}j}e^{i\mathbf{kr}_{\mathbf{n}j}^{\alpha}}d_{\mathbf{n}j\alpha\sigma}^{\phantom{\dag}}\,.
\end{equation}
Here vectors
$\mathbf{r}_{\mathbf{n}j}^{\alpha}
=
\mathbf{r}_{\mathbf{n}}^{\alpha}+j\mathbf{a}_z$
describe positions of sites in the AA~graphite,
${\cal N}$
is the number of elementary unit cells in the three-dimensional (3D)
sample of graphite, and
$\mathbf{k}=(k_x,\,k_y,\,k_z)$
now is a 3D momentum. Its 2D projection,
${\bf k}_{\|}=(k_x, k_y)$,
is confined to the usual hexagonal Brillouin zone of the single-layer
graphene, while
$k_z$
lies in the region
$0<k_z<2\pi/c$.
The Brillouin zone of the AA-G has a shape of right hexagonal prism with
height
$2\pi/c$
(see
Fig.~\ref{FigFermiSurface}).

In terms of spinor~\eqref{eq::Psi} (where now
$\mathbf{k}$
is the 3D vector), Hamiltonian~\eqref{eq::HAAtb} takes the form
\begin{equation}
\label{eq::h_AA}
H^{\rm AA}_{\sigma}=-\sum\limits_{\mathbf{k}}\psi^\dag_{{\bf k}\sigma}
\begin{pmatrix}
2 t_0 \cos(k_z c)& tf (\mathbf{k}_{\|})\\
tf^{*} (\mathbf{k}_{\|})& 2t_0\cos(k_z c)
\end{pmatrix}
\psi^{\phantom{\dag}}_{{\bf k}\sigma}\,.
\end{equation}
This Hamiltonian can be easily diagonalized. The corresponding bands are
\begin{eqnarray}
\varepsilon^{(1)}_{\bf k}&=&-2 t_0 \cos(k_z c)-t |f(\mathbf{k}_{\|})|\,,
\nonumber\\
\varepsilon^{(2)}_{\bf k}&=&-2 t_0 \cos(k_z c)+t |f(\mathbf{k}_{\|})|\,.
\label{eq::tb_bands}
\end{eqnarray}
In a generic situation the Fermi surface of the AA-G consists of two sheets
defined by equations
$\varepsilon^{(1,2)} = \mu$,
where $\mu$ is the chemical potential. In this paper we consider the
undoped compound only. As we will show below this corresponds to the case
$\mu = 0$.
For such a value of $\mu$ the Fermi surface sheets are given by the
relations
\begin{eqnarray}
\varepsilon^{(1)}_{\bf k}=0\;
\Rightarrow\,&&
2t_0\cos(k_{z}c)=-t|f(\mathbf{k}_{\|})|\,,
\label{eq::sheet1}
\\
\varepsilon^{(2)}_{\bf k}=0\;
\Rightarrow\,&&
2t_0\cos(k_{z}c)=t|f(\mathbf{k}_{\|})|\,.
\label{eq::sheet2}
\end{eqnarray}
The AA-G Fermi surface is shown in
Fig.~\ref{FigFermiSurface}. The sheet corresponding to the band
$\varepsilon^{(1)}_{\bf k}$
is the hole sheet, because the component of the velocity vector
$\mathbf{v}_{\mathbf{k}}^{(1)}=\partial\varepsilon^{(1)}_{\bf k}/\partial\mathbf{k}$
normal to the sheet is negative for all momenta on this sheet. Similarly one can prove that the sheet corresponding to the band
$\varepsilon^{(2)}_{\bf k}$
is the electron-like. The states inside the electron (hole) sheet are
filled (empty). Since the sheets have identical volumes, the total number
of electrons in the system per atom is equal to unity. Thus, the case
$\mu=0$,
indeed, corresponds to the undoped AA-G. The Fermi surface of the AA-G has
been studied in several
publications~\cite{aa_graphite_charlier1994,aa_tb_charlier1991,
aa_abinit_charlier1992}.
The results of theses studies are similar to those shown in
Fig.~\ref{FigFermiSurface}.

The surfaces specified by
Eqs.~\eqref{eq::sheet1}
and~\eqref{eq::sheet2}
can be superposed by a parallel translation along
$z$-axis.
Indeed, after transformation
$k_z \rightarrow k_z + \pi/c$
equation~\eqref{eq::sheet1}
becomes
Eq.~\eqref{eq::sheet2},
and vice versa. When a hole Fermi surface sheet may be superposed with an
electron sheet by a suitable translation in momentum space, one refers to
such a Fermi surface as nested. The translation vector superposing the
sheets is called a nesting vector. In our case the nesting vector is
\begin{eqnarray}
\label{eq::nest_vect}
{\bf Q}_0 = \left(0, 0, \frac{\pi}{c} \right).
\end{eqnarray}
The bands
$\varepsilon^{(1)}_{\bf k}$
and
$\varepsilon^{(2)}_{\bf k}$
satisfy the relation
$\varepsilon^{(1)}_{{\bf k}+\mathbf{Q}_0}=-\varepsilon^{(2)}_{\bf k}$.
A Fermi surface with nesting becomes unstable in the presence of arbitrary
weak electron-electron repulsion. Vector
${\bf Q}_0$
characterizes the spatial oscillations of the most unstable mode. The
instability will be discussed in the next section.

\section{Spin-density wave in the AA~graphite}
\label{sec::mean_f}

The instability of the electron liquid with a nested Fermi surface is a
well-known feature. In the majority of papers studying the systems the with
Fermi surface nesting it is accepted that the electron-electron interaction
stabilizes the spin-density wave ground state. Such a picture is used, for
examples, to describe antiferromagnetism in chromium and its
alloys~\cite{Rice,Fawcet_RMP1988_SDW_Cr},
superconducting iron
pnictides\cite{eremin_chub2010,teitel2010,OurPnics2013},
and AA-stacked bilayer
graphene~\cite{PrlOur,PrbROur,PrbOur}.
Minimal model with electron interaction is the Hubbard model. It accounts
for on-site electron-electron interaction only. In this paper we will study
the AA-G version of the Hubbard model  in the framework of the mean-field
approximation. The Hamiltonian of this model is
$H=H^{\text{AA}}+H_{\text{int}}$,
where
$H^{\rm AA}$
is given by
Eq.~\eqref{eq::HAAtbfull}, and
\begin{equation}\label{eq::hubb_H}
H_{\text{int}}=U\sum_{\mathbf{n}i\alpha}\left(n_{\mathbf{n}i\alpha\uparrow}-\frac{1}{2}\right)\left(n_{\mathbf{n}i\alpha\downarrow}-\frac{1}{2}\right).
\end{equation}
Parameter
$U>0$
characterizes on-site electron-electron repulsion, and operator
$n_{\mathbf{n}i\alpha\sigma}=d^\dag_{\mathbf{n}i\alpha\sigma}d^{\phantom{\dag}}_{\mathbf{n}i\alpha\sigma}$.

In SDW state each site acquires a non-zero magnetic moment. We assume here
that all spins are directed parallel or antiparallel to the
$z$-axis.
Thus, the non-zero spin projections are
$S_{\mathbf{n}i\alpha}^{z}
=
(
	\langle n_{\mathbf{n}i\alpha\uparrow}\rangle
	-
	\langle n_{\mathbf{n}i\alpha\downarrow}\rangle
)/2$.
We assume also that the total charge in each site remains constant, that is
$\langle n_{\mathbf{n}i\alpha\uparrow}\rangle
+
\langle n_{\mathbf{n}i\alpha\downarrow}\rangle=1$.
The nesting vector
$\mathbf{Q}_0$
determines the form of the spin-density wave in real space. Specifically,
one can write for the SDW state under study the following equality
\begin{equation}
\label{eq::Sz}
S_{\mathbf{n}j\alpha}^{z}
=
e^{i\mathbf{Q}_0\mathbf{r}_{\mathbf{n}j}^{\alpha}}S_{\alpha}\,.
\end{equation}
Substituting
expression~\eqref{eq::nest_vect}
for
$\mathbf{Q}_0$
into
Eq.~(\ref{eq::Sz}),
one derives
\begin{equation}
\label{eq::layer}
S_{\mathbf{n}j\alpha}^{z}=(-1)^{j}S_{\alpha}\,.
\end{equation}
This shows that spin arrangements in odd and even layers are different from
each other: spin polarizations at two sites separated by vector
${\bf a}_z$
are antiparallel. Yet,
equation~(\ref{eq::layer})
does not specify
$S_{\alpha}$.
Precise structure of
$S_{\alpha}$
has physically relevant consequences. For example, the case
$S_A=S_B$
corresponds to the antiferromagnetically ordered ferromagnetic layers,
while in the case
$S_A=-S_B$
we obtain the so-called G-type antiferromagnetism, where both in-plane and
out-of-plane neighboring spins are antiparallel. One can prove that for the
case
$S_A=S_B$
the gap at the Fermi level does not arise and this state is unstable. At
the same time, the SDW state with
$S_A=-S_B$
does open the gap at the Fermi level for arbitrary small $U$, and
corresponds to the mean-field ground state of the
model~\eqref{eq::hubb_H}.
Spin configuration for this SDW order is shown in
Fig.~\ref{ris::aa_latt}.

To describe such an ordered state we introduce the SDW order parameter
\begin{equation}
\label{Deltaia}
\Delta_{i\alpha}
=
\frac{U}{2}
\left( \bar{n}_{i\alpha\uparrow} - \bar{n}_{i\alpha\downarrow}\right).
\end{equation}
Here
$\bar{n}_{i\alpha\sigma}=\langle n_{\mathbf{n}i\alpha\sigma}\rangle$.
The order parameter satisfies the conditions
\begin{equation}
\label{eq::sdw_mag}
\Delta_{i\alpha}
=
(-1)^i\Delta_{\alpha}\,,\;\;\Delta_{A}=-\Delta_{B}\equiv\Delta\,.
\end{equation}
In mean-field approximation, we decompose the density operator in
Eq.~\eqref{eq::hubb_H}
as follows
$n_{\mathbf{n}i\alpha\sigma}
=
\bar{n}_{i\alpha\sigma}+\delta n_{\mathbf{n}i\alpha\sigma}$,
where operators
$\delta n_{\mathbf{n}i\alpha\sigma}
=
n_{\mathbf{n}i\alpha\sigma}- \bar{n}_{i\alpha\sigma}$
describe fluctuations near the average density
$\bar{n}_{i\alpha\sigma}$.
Mean-field interaction Hamiltonian is obtained by neglecting the terms
quadratic in
$\delta n_{\mathbf{n}i\alpha\sigma}$.
As a result, we derive
\begin{equation}
\label{UMF}
H_{\text{int}}^{\rm MF}
=
\sum_{\mathbf{n}i\alpha}
	\left[
		-\Delta_{i\alpha}
		\left(
			n_{\mathbf{n}i\alpha\uparrow}
			-
			n_{\mathbf{n}i\alpha\downarrow}
		\right)
		+
		\frac{\Delta_{i\alpha}^2}{U}
	\right].
\end{equation}

The considered SDW state doubles the lattice period in the
$z$-direction,
while preserving the translation invariance along the layers. Consequently,
the elementary cell in the ordered phase contains four sites, two sites in
one layer and two sites in an adjacent layer. Due to the doubling of the
elementary cell, the Brillouin zone shrinks in
$k_z$-direction:
now, projection
$k_z$
varies from $0$ to
$\pi/c$.
For further analysis it is convenient to introduce the following
$4$-component
spinor:
\begin{equation}
\Psi_{\mathbf{k}\sigma}^{\phantom{\dag}}
=
\sqrt{\frac{2}{\cal N}}
\sum_{\mathbf{n}j}e^{i\mathbf{kr}_{\mathbf{n}2j}^{\alpha}}
\begin{pmatrix}
d_{\mathbf{n}2jA\sigma}^{\phantom{\dag}}\\
d_{\mathbf{n}2jB\sigma}^{\phantom{\dag}}\\
d_{\mathbf{n}2j+1A\sigma}^{\phantom{\dag}}\\
d_{\mathbf{n}2j+1B\sigma}^{\phantom{\dag}}
\end{pmatrix}.
\end{equation}
In terms of this spinor, the total mean-field Hamiltonian can be written as
\begin{equation}
H^{\rm MF}
=
2{\cal{N}}\frac{\Delta^2}{U}
+
\mathop{{\sum}'}_{\mathbf{k}\sigma}\!
	\Psi_{\mathbf{k}\sigma}^{\dag}
		\hat{H}^{\rm MF}_{\mathbf{k}\sigma}
	\Psi_{\mathbf{k}\sigma}^{\phantom{\dag}}\,,
\end{equation}
where the summation symbol with prime denotes the summation over the
reduced Brillouin zone, and
$4\times4$
matrix
$\hat{H}^{\rm MF}_{\mathbf{k}\sigma}$
equals to
\begin{equation}
\hat{H}^{\rm MF}_{\mathbf{k}\sigma}
=
-
\begin{pmatrix}
 \Delta_\sigma           & tf(\mathbf{k}_{\|})& \!\!\!t_0 g (k_z) &0                  \\
tf^*(\mathbf{k}_{\|}) &   -\Delta_\sigma   &        0
&\!\!\!t_0 g (k_z) \\
t_0 g^* (k_z)           &   0                &    -\Delta_\sigma         &tf(\mathbf{k}_{\|})\\
0                       &\!\!\!t_0 g^* (k_z) &   tf^*(\mathbf{k}_{\|}) & \Delta_\sigma
\end{pmatrix}\!.
\end{equation}
Here
$\Delta_{\uparrow}=\Delta$,
and
$\Delta_{\downarrow}=-\Delta$,
and function $g$ is defined as
$g(k_z) = 1 + e^{2ik_zc}$.
Matrix
$\hat{H}^{\rm MF}_{\mathbf{k}\sigma}$
can be easily diagonalized. The mean-field eigenenergies are independent of
electron spin and equal to
\begin{eqnarray}
E_\mathbf{k}^{(1)}&=&-\sqrt{\Delta^2+\left[t|f(\mathbf{k}_{\|})|+2t_0\cos(c k_z)\right]^2}\,,\label{eq::MF_band1}\\
E_\mathbf{k}^{(2)}&=&-\sqrt{\Delta^2+\left[t|f(\mathbf{k}_{\|})|-2t_0\cos(c k_z)\right]^2}\,,\label{eq::MF_band2}\\
E_\mathbf{k}^{(3)}&=&\sqrt{\Delta^2+\left[t|f(\mathbf{k}_{\|})|-2t_0\cos(c k_z)\right]^2}\,,\label{eq::MF_band3}\\
E_\mathbf{k}^{(4)}&=&\sqrt{\Delta^2+\left[t|f(\mathbf{k}_{\|})|+2t_0\cos(c k_z)\right]^2}\,.\label{eq::MF_band4}
\end{eqnarray}
At half-filling and zero temperature
$T=0$
first two bands are filled, last two are empty, and the system is an
insulator with the gap equal to $2\Delta$. Consequently, the
zero-temperature mean-field energy is
\begin{eqnarray}
E_{\rm MF}=2{\cal{N}}\frac{\Delta^2}{U}+2\mathop{{\sum}'}_{\mathbf{k}}\!\left(E_\mathbf{k}^{(1)}+E_\mathbf{k}^{(2)}\right).
\end{eqnarray}

Self-consistent equation for the order parameter is obtained by
minimization of
$E_{\rm MF}$
with respect to $\Delta$. Taking into account that
$\mathop{{\sum}'}_{\mathbf{k}}
[E_\mathbf{k}^{(1)}
+
E_\mathbf{k}^{(2)}]
\!=\!
\sum_{\mathbf{k}}\!E_\mathbf{k}^{(2)}$,
where the summation on the right-hand side is performed over the full AA-G
Brillouin zone, we can write the self-consistency equation
$\partial E_{\rm MF}/\partial\Delta=0$
as
\begin{equation}
\label{eq::diff_E_MF}
\frac{2}{U}
=
\int\limits_{-\infty}^{\infty}\!\!
	d\varepsilon\,
	\frac{\rho (\varepsilon) }{\sqrt{\Delta^2+\varepsilon^2}}\,.
\end{equation}
In this equation the AA-G density of states
$\rho(\varepsilon)$
is defined according to the formula
\begin{equation}
\label{eq::dens_defin}
\rho (\varepsilon)
=
\int\frac{d^3{\bf k}}{v_{\rm BZ}}\,
	\delta\!\left(
			t|f_{\mathbf{k}_{\|}}|
			+
			2t_0 \cos (k_zc)
			-
			\varepsilon
		\right),
\end{equation}
in which the integration is performed over the full AA-G Brillouin zone,
and
$v_{\rm BZ}=16\pi^3/(3\sqrt{3}c a^2)$
is the Brillouin zone volume. Since for any
$\mathbf{k}_{\|}$
one has
$0< |f({\mathbf{k}_{\|}})| < 3$,
the density of
states~\eqref{eq::dens_defin}
is non-zero in the range
$-2t_0<\varepsilon<3t+2t_0$.

It is convenient to express the density of states,
Eq.~\eqref{eq::dens_defin},
as a sum
\begin{equation}
\label{eq::total_density}
\rho(\varepsilon)
=
\rho_{\rm gr}(\varepsilon)\Theta(\varepsilon)+\delta\rho(\varepsilon)\,,
\end{equation}
where
$\rho_{\rm gr}(\varepsilon)$
is the density of states of the single layer graphene,
$\Theta(\varepsilon)$
is the Heaviside step function, and correction
$\delta \rho (\varepsilon)$
vanishes when
$t_0 = 0$.
The term
$\delta \rho$
corresponds to modification of the density of states due to the inter-layer
hopping. In the (realistic) limit
$t_0 \ll t$
and for small energy
$\varepsilon\ll t$
the following approximate expression for
$\delta \rho$
may be established (see
Appendix~\ref{app::dos})
\begin{eqnarray}
\label{eq::density_d}
\delta\rho(\varepsilon)\!\!
&\approx&\!\!
\frac{2}{\sqrt{3} \pi^2 t^2}
\Theta(2t_0 - |\varepsilon|)\times
\\
\nonumber
&&\left[
	\sqrt{4t_0^{2} - \varepsilon^2}
	-
	|\varepsilon| \arccos \left(\dfrac{|\varepsilon|}{2t_0}\right)
\right].
\nonumber
\end{eqnarray}
Formally, this expression was derived in the low-energy limit
$\varepsilon \ll t$.
Fortunately,
decomposition~(\ref{eq::total_density})
with
$\delta \rho$
given by
Eq.~(\ref{eq::density_d})
works quite well almost everywhere, except near the van~Hove singularity
$\varepsilon \sim t$,
and the band edge
$\varepsilon \sim 3t$,
see discussion in
Appendix~\ref{app::dos}
and
Fig.~\ref{FigDOS}.
\begin{figure}[t]
\center{\includegraphics[width=0.95\columnwidth]{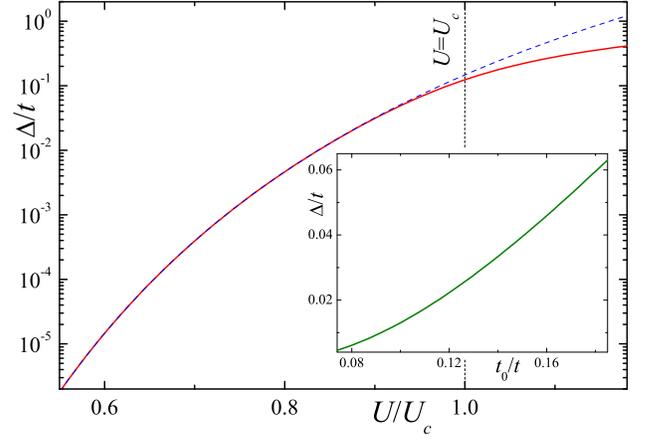}}
\caption{The dependence of the SDW order parameter on the on-site repulsion
energy $U$. The plots are calculated for
$t_0=0.37$\,eV
and
$t=2.7$\,eV,
which corresponds to the ratio
$t_0/t=0.136$.
Red solid curve is found numerically by solving
Eq.~\eqref{eq::diff_E_MF},
while dashed blue curve corresponds to approximate
formula~\eqref{eq::M}.
Inset shows the dependence of $\Delta$ on
$t_0$
calculated by solving
Eq.~\eqref{eq::diff_E_MF}
at
$U/U_c=0.9$.
\label{FigDeltaU}}
\end{figure}

Equations~(\ref{eq::total_density})
and~(\ref{eq::density_d})
allow one to estimate the integral in
Eq.~\eqref{eq::diff_E_MF}
and obtain an analytical expression for the SDW order parameter in the
limit
$\Delta\ll t$.
To this end we re-write
Eq.~\eqref{eq::diff_E_MF}
in the following manner:
\begin{equation}
\label{eq::diff_E_MF1}
\frac{2}{U}=\int\limits_{0}^{3t}\!\!d\varepsilon\,
	\frac{\rho_{\text{gr}}(\varepsilon)}{\sqrt{\Delta^2+\varepsilon^2}}
+\!\!\!
\int\limits_{-2t_0}^{2t_0}\!\!\!d\varepsilon\,
	\frac{\delta\rho(\varepsilon)}{\sqrt{\Delta^2+\varepsilon^2}}\,.
\end{equation}
Since
$\rho_{\rm gr}(\varepsilon)\propto\varepsilon$
at small energies, the first integral in this formula is well-defined for
$\Delta\to0$.
It equals
\begin{equation}
\label{eq::int1}
\int\limits_{0}^{3t}\!\!d\varepsilon\,\frac{\rho_{\text{gr}}(\varepsilon)}{\sqrt{\Delta^2+\varepsilon^2}}
\approx\!\int\limits_{0}^{3t}\!\!d\varepsilon\,\frac{\rho_{\text{gr}}(\varepsilon)}{\varepsilon}\equiv\frac{2}{U_c}\,.
\end{equation}
Constant
$U_c$,
defined by this equation, has the dimension of energy. Its physical meaning
will be described below. Numerical calculations of the
integral~(\ref{eq::int1})
with full density of states of graphene give
$U_c=2.23t$.
For
$t=2.7$\,eV, we have
$U_c\approx6.02$\,eV.

The second integral in
Eq.~(\ref{eq::diff_E_MF1})
diverges logarithmically when $\Delta$ vanishes. It requires a more
cautious approach. The detailed calculations are relegated to
Appendix~\ref{app::selfconsist}.
The resultant expression is
\begin{equation}
\label{eq::int2}
\int\limits_{-2t_0}^{2t_0}\!\!\!d\varepsilon\,
	\frac{\delta\rho(\varepsilon)}{\sqrt{\Delta^2+\varepsilon^2}}
\approx
2\rho_0\left(\ln\frac{8t_0}{\Delta}-2\right),
\end{equation}
where the AA-G density of states at the Fermi level equals
\begin{eqnarray}
\label{eq::rho0}
\rho_0=\rho(0)
=
\delta \rho (0)
\approx
\frac{4t_0}{\sqrt{3}\pi^2 t^2}.
\end{eqnarray}
Combining
Eqs.~\eqref{eq::diff_E_MF1},
\eqref{eq::int1},
and~\eqref{eq::int2},
we derive the following relation for the SDW order parameter:
\begin{equation}\label{eq::M}
\Delta
\approx
8t_0\exp\left[-\frac{1}{\rho_0}\left(\frac{1}{U}-\frac{1}{U_c}\right)-2
       \right].
\end{equation}
This equation is valid for small $\Delta$. As the gap grows, this
analytical expression becomes progressively less accurate. In such a
situation, one is forced to solve
Eq.~\eqref{eq::diff_E_MF}
numerically. The dependence of $\Delta$ on $U$ calculated numerically and
estimated according
approximation~\eqref{eq::M},
are shown in
Fig.~\ref{FigDeltaU}.
The data in the figure demonstrate an excellent agreement between the two
approaches if
$U < U_c$.

\section{Discussion}\label{sec::conclusions}

\subsection{Single-layer graphene physics in SDW transition}

Our theory implies that the AA-G is a SDW insulator at low temperature. The
value of the insulating gap
$2\Delta$
substantially depends on the interaction parameter $U$ and the inter-layer
hopping amplitude
$t_0$
[see inset to
Fig.~\ref{FigDeltaU}
and
Eq.~(\ref{eq::M})].
The sensitivity to $U$ is a familiar feature of a mean field theory. As for
the dependence on 
$t_0$,
it is a consequence of the fact that the AA-G density of states at the
Fermi level
$\rho_0$
is proportional to
$t_0$.
Reducing
$t_0$
to zero, we enter a regime where our model describes a collection of
decoupled graphene layers. Due to its importance, let us analyze this
limit in more detail.

Equation~\eqref{eq::M}
implies that
$\Delta\to0$
when
$t_0\to0$,
provided that $U$ is smaller than the critical threshold
$U_c$.
For
$U>U_c$,
equation~\eqref{eq::M}
predicts that $\Delta$ diverges when
$t_0\to0$,
indicating the failure of
approximation~\eqref{eq::M}
for large $U$. The value
$U_c \approx 2.23 t \approx6.02$\,eV
is found using
Eq.~\eqref{eq::int1}.
It can be also calculated from
Eq.~\eqref{eq::diff_E_MF}
in the limit
$\Delta=t_0=0$.

The difference between
$U < U_c$
and
$U > U_c$
regimes is physically significant. Once
$U>U_c$,
Eq.~\eqref{eq::diff_E_MF}
has a solution even for uncoupled layers, when
$t_0=0$.
In other words, the ground state of the Hubbard model for single graphene
layer is antiferromagnetic for
$U>U_c$.
This is a well-known
result~\cite{GrapheneHubbardMC,GrapheneHubbardMC_MF,GrapheneHubbardMF}.
The experiments show that graphene remains semimetal even at low
temperatures. Thus, we expect that
$U<U_c$.
The approach exploring Monte-Carlo
simulations~\cite{GrapheneHubbardMC,GrapheneHubbardMC_MF}
gives
$U_c/t\approx4.5$
(or $U_c\approx12.15$\,eV for $t=2.7$\,eV),
which is larger than the the presented above mean-field result
$U_c \approx 2.23 t \approx6.02$\,eV~\cite{GrapheneHubbardMF}.
{\it Ab initio} calculations of the Hubbard $U$ in graphene performed in
Ref.~\onlinecite{Wehling},
give
$U\approx9.3$\,eV,
that is, the value close, but somewhat smaller than critical value
$U_c$
obtained by Monte-Carlo simulations. While the single-layer graphene
physics cannot generate the ordering transition for
$U<U_c$,
it affects the magnitude of $\Delta$ significantly: large factor
$\exp(1/(\rho_0U_c))$
in
Eq.~\eqref{eq::M}
introduces strong renormalization of pre-exponential energy scale
$t_0$.

\subsection{Comparison with AA bilayer graphene}

The presented theory of the SDW order in AA~graphite is an extension of SDW
theory for the AA~bilayer graphene, whose lattice has similar geometric
structure. For the SDW order in AA~bilayer graphene, the mean-field
calculations have been reported in
Refs.~\onlinecite{PrlOur,PrbOur,PrbVOur,PrbROur,BreyFertig},
the investigations by numerical methods have been presented in
Refs.~\onlinecite{Honerkamp,ulybyshev_conf,ulybyshev,num_rev_buivid_ulyb2016}.
These results, as well as some others, were reviewed in
Ref.~\onlinecite{bilayer_review2016}.

Experimental data for AA~graphene are quite limited. This is a consequence
of small number of samples. If one is interested in possible SDW in
AA~bilayer graphene, there is additional experimental complication. The
bilayer, being true 2D material, contains too little amount of matter for a
currently extant neutron scattering techniques to be of use. On the other
hand, AA~graphite is 3D system. Therefore, synthesis of sufficiently bulky
AA~graphite samples may bear significant implications for understanding of
possible magnetism of the AA~bilayer graphene.

\subsection{Other types of order parameters}

As it follows from
Eqs.~\eqref{eq::Sz}
and~(\ref{eq::layer}),
the induced magnetization oscillates in space with the nesting wave vector
${\bf Q}_0$.
This spatial modulation is an important feature for it guarantees the
coupling of the two nested Fermi surface sheets, leading to the SDW
instability. There are other order parameters, which oscillate in space
with
${\bf Q}_0$.
One of them was already mentioned above. It is the order parameter of the
SDW type, with magnetization described by
Eq.~\eqref{eq::Sz}
in which
$S_{\alpha}$
is chosen according to
$S_A=S_B$.
This order corresponds to layered antiferromagnetic state. While it
oscillates with the required wave vector
${\bf Q}_0$,
it does not open a gap at Fermi level, and only modifies the Fermi surface.
This can be easily shown performing calculations similar to the presented
in previous Section. As a result, such an order cannot benefit from
nesting. Similar argumentation was used in
Ref.~\onlinecite{PrlOur}
for the AA bilayer graphene.

Another possible order parameter oscillating with wave vector
$\mathbf{Q}_0$
describes the CDW state. It can be written as
\begin{equation}\label{DeltaniaCDW}
\Delta^{\text{CDW}}_{\mathbf{n}j\alpha}=\frac{U}{2}\left(\langle n_{\mathbf{n}j\alpha\uparrow}\rangle+\langle n_{\mathbf{n}j\alpha\downarrow}\rangle\right)
=e^{i\mathbf{Q}_0\mathbf{r}_{\mathbf{n}j}^{\alpha}}\Delta^{\text{CDW}}_{\alpha}\,.
\end{equation}
Similar to
Eq.~(\ref{eq::sdw_mag}),
the gap at Fermi level is opened, when
$\Delta^{\text{CDW}}_{A}=-\Delta^{\text{CDW}}_{B}$.
However, in our model, the CDW is stable only if
$U<0$,
otherwise, such an order parameter is absolutely unstable. (In principle,
even in repulsive models the CDW can be induced by a sufficiently strong
magnetic
field~\cite{cdw_magfield_theor1981,cdw_magfield_exp2017,
cdw_magfield_exp2001},
or lattice
participation~\cite{cdw_mazin2008}.
However, studying these factors is beyond the present discussion.)

\subsection{Denesting}

It is important to discuss the effects of the violation of perfect nesting
in our model. Analyzing
Eq.~\eqref{eq::M},
we notice that $\Delta$ vanishes exponentially for vanishing interaction
$U$, however, it remains finite for any finite $U$. In this respect our
calculations are very similar to the BCS result for superconducting order
parameter. This feature is a consequence of the perfect nesting of the
Fermi surface sheets. The perfect nesting is an approximation. It may be
destroyed by longer-range hopping processes in the kinetic energy term. For
a Fermi surface with an imperfect nesting the interaction parameter $U$
must exceed some critical strength
$U^*$
to induce the ordering
transition~\cite{OurPnics2013}.
The value of
$U^*$
depends on a degree of the denesting. Therefore, sufficiently strong
denesting prevents SDW order by pushing
$U^*$
above $U$.

In addition to the longer-range hopping amplitudes, the denesting may be
enhanced by doping: extra electrons ``inflate" the electron Fermi surface
sheet and ``deflate" the hole sheet. The hole doping exerts the opposite
effect on the sheets. Regardless of the sign of the doped charge, the
shapes of the sheets become unequal after the doping, violating the nesting.
Doping-induced denesting destabilizes the homogeneous state of the electron
liquid. Theoretical studies of the inhomogeneous states (``stripes", phase
separation) were performed for a variety of
systems~\cite{zaanen_stripes1989,tokatly1992,fflo4,spinstate2009,teitel2010,
IrkhinPRB2010,We_grA_grE2012,PrbOur,WeImperf,bianconi2015intrinsic,prb_sl2017,
half_met_prl2017,rakhmanov2017inhomogeneous}.
It follows from this research that doped SDW systems have rich phase
diagram and demonstrate interesting physical phenomena. Therefore, doped
AA~graphite might deserve a special investigation.

\subsection{Motivation for the use of the Hubbard Hamiltonian}

It is well-known that the use of the Hubbard model, with its extremely
short-range interaction, may be partially justified in case of metals with
short screening length. Unfortunately, the screening in AA~graphite, as
well as in graphene, bilayer graphene, and related materials is rather poor
due to vanishing or low density of states at the Fermi energy.

For AA~graphite, a possible alternative to the Hubbard interaction is the
use of the screened Coulomb interaction consistent with small, but finite,
number of the charge carriers. However, we believe that at the present
phase of the research the use of the Hubbard model is warranted. First of
all, one must remember that the SDW instability in our model is
nesting-driven. Consequently, at the qualitative level, the SDW is fairly
insensitive to details of the interaction. Furthermore, the mean-field
calculations for the Hubbard Hamiltonian are simple and well-understood.
This assures that mathematical details of the formalism will not obstruct
the qualitative discussion. A more rigorous and complex analysis could be
executed at later stages.

Currently, the Hubbard model is a common approach employed for description
of graphene and related
materials~\cite{Nilsson2006,Dillenschneider2008,
lang_af_hubb2012,af_ab_hubb2013,sun_hub_mc_afm2014}.
The ability of the Hubbard interaction to mimic properties of the
longer-range interaction is also
discussed~\cite{EffCoulomb}.
Thus, it appears that, while not without its flaws, the Hubbard Hamiltonian
is a suitable tool for the task at hand.

\subsection{Conclusions}

In this paper we have studied SDW order in AA~graphite. Unlike the
single-layer graphene, whose Fermi surface shrinks to two Fermi points, AA
graphite has a well-developed two-sheet Fermi surface. This Fermi surface
is a consequence of interlayer tunneling, and it disappears when the
tunneling vanishes. The SDW instability is driven by the nesting of two
Fermi surface sheets. Straightforward mean-field calculations allow one to
estimate the SDW order parameter magnitude. The derived expression for the
SDW magnetization shows strong enhancement due to single-layer-graphene
electron states.

\section*{Acknowledgments}  This work is partially supported by the Russian
Foundation for Basic Research (Projects 17-02-00323).

\appendix

\section{Calculation of density of states}
\label{app::dos}

In this Appendix we calculate density of states
$\rho(\varepsilon )$,
which is defined by
Eq.~\eqref{eq::dens_defin}.
In general, the argument of the $\delta$-function in the integral of
Eq.~\eqref{eq::dens_defin}
is complicated. However, in the limit
$\varepsilon\ll t$
and
$t_0\ll t$
one can replace
$t|f_{\mathbf{k}_{\|}}|\approx v_{\rm F}|\mathbf{q}|$.
In this regime, we evaluate the integral in
Eq.~\eqref{eq::dens_defin}
explicitly
\begin{eqnarray}
\rho(\varepsilon)
&=&
\int\frac{d^3\mathbf{k}}{v_{\rm BZ}}\,
	\delta\!\left(
			t|f_{\mathbf{k}_{\|}}|
			+
			2t_0 \cos (k_zc)
			-
			\varepsilon
		\right)
\\
&\approx&
N_{\rm D}\int\frac{d^2\mathbf{q}dk_z}{v_{\rm BZ}}\,
	\delta\!\left(
			v_{\rm F}|\textbf{q}|
			+
			2t_0\cos(k_z c)
			-
			\varepsilon
			\right)
\nonumber
\\
&=&
\frac{4\pi}{v_{\rm BZ}}\!\!
\int\limits_{0}^{2\pi /c}\!\!\!dk_z\!\!
	\int\limits_{0}^{\infty}\!qdq\,
		\delta\left(v_{\rm F}q+2t_0\cos(k_z c)-\varepsilon\right)
\nonumber\\
&=&
\frac{4\pi}{c v_{\rm F}^2v_{\rm BZ}^{\vphantom{2}}}
\int\limits_{0}^{2\pi }\!\!d\gamma\,
	(\varepsilon-2t_0\cos\gamma)
	\Theta(\varepsilon-2t_0\cos\gamma)\,.
\nonumber
\end{eqnarray}
Symbol
$N_{\rm D}=2$
denotes the number of non-equivalent Dirac points, and
$\Theta(x)$
is the Heaviside step function. Taking into account that
$v_{\text{F}}=3ta/2$
and
$v_{\rm BZ}=16\pi^3/(3\sqrt{3}ca^2)$,
the density of states
$\rho(\varepsilon)$
can be expressed as
\begin{equation}\label{eq::rho_F}
\rho(\varepsilon)=\frac{2t_0}{\sqrt{3}\pi^2t^2}F(\varepsilon/2t_0)\,,
\end{equation}
where dimensionless function
$F(\xi)$
is equal to
\begin{eqnarray}
F(\xi)
=
\int\limits_{0}^{2\pi}\!\!
	d\gamma(\xi-\cos\gamma)\Theta(\xi-\cos\gamma)
=
2\pi\xi\Theta(\xi)
\nonumber
\\
+\left(2\sqrt{1-\xi^2}+2\xi \arcsin \xi-\pi|\xi|\right)\Theta(1-|\xi|)\,.
\end{eqnarray}
Combining the latter equation with
Eq.~\eqref{eq::rho_F},
we determine
\begin{eqnarray}
\rho(\varepsilon)
&=&
\frac{2\varepsilon}{\sqrt{3}\pi t^2}\,\Theta(\varepsilon)+
\frac{2}{\sqrt{3}\pi^2t^2}\Theta\left(2t_0-|\varepsilon| \right)\times\nonumber\\
&&\left[\sqrt{4t_0^{2}-\varepsilon^2}-|\varepsilon| \arccos \left(\frac{|\varepsilon|}{2t_0}\right)\right]\,.\label{eq::DOSappr}
\end{eqnarray}
The first term in this equation corresponds to the well-known low-energy
approximation for the density of states of the single layer graphene:
\begin{equation}
\label{eq::grDOS0}
\rho_{\rm gr}(\varepsilon)\approx\frac{2|\varepsilon|}{\sqrt{3}\pi t^2}\,.
\end{equation}
The second term, which is equal to
$\delta\rho(\varepsilon)$
from
Eq.~\eqref{eq::density_d},
is the correction due to the inter-layer tunneling. This correction is of
the order of
$t_0/t$.
It is non-zero only for
$|\varepsilon|<2t_0$.

\begin{figure}[t]
\center{\includegraphics[width=0.95\columnwidth]{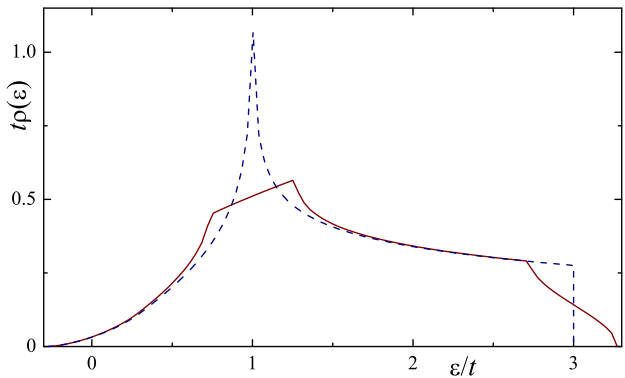}}
\caption{The density of states of the AA~graphite versus energy. The plots
are calculated for $t_0=0.37$\,eV and $t=2.7$\,eV,
which corresponds to the ratio
$t_0/t=0.136$.
Solid red curve is the result of numerical computation of integral in
Eq.~\eqref{eq::dens_defin}.
Blue dashed curve corresponds to
formula~\eqref{eq::total_density}
in which approximate
expression~(\ref{eq::density_d})
for
$\delta \rho$
was used.
\label{FigDOS}}
\end{figure}
When the condition
$\varepsilon\ll t$
is violated,
Eq.~\eqref{eq::DOSappr}
is no longer valid, and more elaborate approach is necessary. Integrating
over
${\bf k}_\| = (k_x, k_y)$
in
Eq.~\eqref{eq::dens_defin}
one derives
\begin{equation}
\label{eq::DOSAAG2gr}
\rho(\varepsilon)
=\!
\int\limits_{0}^{2\pi}\!\frac{d\gamma}{2\pi}\,
	\rho_{\text{gr}}\!
	\left(\varepsilon-2t_0\cos \gamma \right)
	\Theta \left(\varepsilon-2t_0\cos \gamma \right),
\end{equation}
This integral can be evaluated numerically, using, for example, numerically
exact graphene density of state
$\rho_{\text{gr}}(\varepsilon)$.
As a result, one accurately obtains the density of states for
Hamiltonian~(\ref{eq::HAAtbfull}).
However, for our mean-field treatment a less rigorous form of
$\rho (\varepsilon)$
is acceptable: we can employ
decomposition~(\ref{eq::total_density})
with
$\delta \rho$
given by the approximate
expression~(\ref{eq::density_d}).
Figure~\ref{FigDOS}
attests to the quality of this approximation. We see that both functions
are virtually identical except the energies near the van~Hove singularity
$\varepsilon = t$
and the high-energy band edge
$\varepsilon = 3t$.
Such a success may be explained as follows. Expanding
Eq.~(\ref{eq::DOSAAG2gr})
in powers of
$t_0$,
one writes
\begin{equation}
\label{eq::dos_taylor}
\rho(\varepsilon)
\approx
\rho_{\text{gr}}(\varepsilon)
+
t_0^2\rho_{\text{gr}}''(\varepsilon)\,.
\end{equation}
This expression is valid away from the van~Hove singularity and spectrum
edges, where function
$\rho_{\text{gr}}(\varepsilon) \Theta (\varepsilon)$
does not have well-defined derivatives. In
Eq.~(\ref{eq::dos_taylor})
the correction of the order of
$t_0$
is zero. Neglecting small terms of the order of
$t_0^2$,
we conclude that, away from the points
$\varepsilon = 0$,
$\varepsilon = t$,
and
$\varepsilon = 3t$,
we can approximate
$\rho(\varepsilon)\approx\rho_{\text{gr}}(\varepsilon)$.
Taking into account the low-energy correction
$\delta \rho (\varepsilon)$,
Eq.~(\ref{eq::density_d}),
we capture the behavior of the density of states near
$\varepsilon = 0$.
Quality of approximation remains poor near
$\varepsilon = t$
and
$\varepsilon = 3t$.
These regions, fortunately, contribute weakly to the mean-field properties
of the model. Thus, we accept that
Eqs.~\eqref{eq::total_density}
and~(\ref{eq::density_d})
give a very good approximation to the AA-G density of states.

\section{Evaluation of the self-consistency equation}
\label{app::selfconsist}

In this Appendix we will evaluate the integral presented in
Eq.~\eqref{eq::int2}.
It diverges when
$\Delta\to0$.
To evaluate this integral the divergent term must be treated separately
from the finite contribution. To this end we write
\begin{equation}\label{eq::split_integral}
\int\limits_{-2t_0}^{2t_0}\!\!\!d\varepsilon\,\frac{\delta\rho(\varepsilon)}{\sqrt{\Delta^2+\varepsilon^2}}=I_1+I_2\,,
\end{equation}
where the quantities
$I_{1,2}$
are defined by the following relations
\begin{eqnarray}
I_1
&=&\!\!\!\!
\int\limits_{-2t_0}^{2t_0}\!\!\!d\varepsilon\,
	\frac{
		\delta\rho(\varepsilon)-\delta\rho(0)
	     }
	     {
		\sqrt{\Delta^2+\varepsilon^2}
	     }\,,
\\
I_2
&=&\!\!\!\!
\int\limits_{-2t_0}^{2t_0}\!\!\!d\varepsilon\,
	\frac{\delta\rho(0)}{\sqrt{\Delta^2+\varepsilon^2}}
=
2\rho_0\arsinh\left(\frac{2t_0}{\Delta}\right).
\end{eqnarray}
Symbol
$\rho_0$
is defined by
Eq.~(\ref{eq::rho0}).
For small $\Delta$ one has
\begin{equation}\label{A2::I2}
I_2\approx2\rho_0\ln\left(\dfrac{4t_0}{\Delta}\right).
\end{equation}
Integral
$I_1$
remains finite when
$\Delta\to0$
and can be approximated by its value at
$\Delta=0$:
\begin{equation}
I_1\approx\frac{8t_0}{\sqrt{3}\pi^2t^2}\!\int\limits_{0}^{2t_0}\!\frac{d\varepsilon}{\varepsilon}\left[\sqrt{1-\frac{\varepsilon^2}{4t^2_0}}
-\frac{\varepsilon}{2t_0}\arccos\!\left(\!\dfrac{\varepsilon}{2t_0}\!\right)-1\right].
\end{equation}
Since
\begin{eqnarray}
\int\limits_{0}^{2t_0}\!\frac{d\varepsilon}{2t_0}\arccos \left(\frac{\varepsilon}{2t_0}\right)&=&1\,,\\
\int\limits_{0}^{2t_0}\!\frac{d\varepsilon}{\varepsilon}\left[\sqrt{1-\frac{\varepsilon^2}{4t^2_0}}-1	 \right]&=&\ln 2 -1\,,
\end{eqnarray}
we can estimate
$I_1$
as follows
\begin{equation}
I_1\approx\frac{8t_0}{\sqrt{3}\pi^2t^2}(\ln2-2)=2\rho_0\,(\ln2-2)\,.
\end{equation}
Combining this expression with
Eq.~\eqref{A2::I2},
one obtains
\begin{eqnarray}
\int\limits_{-2t_0}^{2t_0}\!\!\!d\varepsilon\,
	\frac{\delta\rho(\varepsilon)}{\sqrt{\Delta^2+\varepsilon^2}}
\approx
2\rho_0\,\left(\ln\frac{8t_0}{\Delta}-2\right).
\end{eqnarray}
This concludes the derivation of
Eq.~\eqref{eq::int2}.


\end{document}